\documentclass[letterpaper]{article} 
\usepackage{generic}
\usepackage{cite}
\usepackage{amssymb,amsfonts}
\usepackage{graphicx}

\usepackage{amsmath}
\usepackage{amsthm}
\usepackage{booktabs}

\usepackage{textcomp}
\usepackage{multirow}
\usepackage{tabularx}

\newtheorem{definition}{Definition}

\usepackage{aaai2026}  
\usepackage{times}  
\usepackage{helvet}  
\usepackage{courier}  
\usepackage[hyphens]{url}  
\usepackage{graphicx} 
\usepackage{natbib}  
\usepackage{caption} 
\usepackage{algorithm}
\usepackage{algorithmic}

\usepackage{newfloat}
\usepackage{listings}
\DeclareCaptionStyle{ruled}{labelfont=normalfont,labelsep=colon,strut=off} 
\floatstyle{ruled}
\newfloat{listing}{tb}{lst}{}
\floatname{listing}{Listing}
\title{Experience with Single Domain Generalization in Real World Medical Imaging Deployments}
\author{
    Ayan Banerjee\textsuperscript{\rm 1},  Komandoor Srivathsan\textsuperscript{\rm 2}, Sandeep K.S. Gupta\textsuperscript{\rm 1}\\
}
\affiliations{
    \textsuperscript{\rm 1}Arizona State University,\\ \textsuperscript{\rm 2}Mayo Clinic\\

    \{abanerj3, sandeep.gupta\}@asu.edu, srivathsan.komandoor@mayo.edu

}

\usepackage{bibentry}
\begin{document}

\maketitle

\begin{abstract}
A desirable property of any deployed artificial intelligence is generalization across domains, i.e. data generation distribution under a specific acquisition condition. In medical imagining applications the most coveted property for effective deployment is  Single Domain Generalization (SDG), which addresses the challenge of training a model on a single domain to ensure it generalizes well to unseen target domains.  In multi-center studies, differences in scanners and imaging protocols introduce domain shifts that exacerbate variability in rare class characteristics. This paper presents our experience on SDG in real life deployment for two exemplary medical imaging case studies on seizure onset zone detection using fMRI data, and stress electrocardiogram based coronary artery detection. Utilizing the commonly used application of diabetic retinopathy, we first demonstrate that state-of-the-art SDG techniques fail to achieve generalized performance across data domains. We then develop a generic expert knowledge integrated deep learning technique DL+EKE and instantiate it for the DR application and show that DL+EKE outperforms SOTA SDG methods on DR. We then deploy instances of DL+EKE technique on the two real world examples of stress ECG and resting state (rs)-fMRI and discuss issues faced with SDG techniques.
\end{abstract}


\section{Introduction}
\label{sec:introduction}
Domain generalization characterizes a model’s capacity to maintain performance when exposed to data that diverges from the distribution it encountered during training~\cite{stolte2023domino}. This capability is especially vital in medical AI, where models must reliably interpret previously unseen patient cases originating from different clinical environments. In medical imaging, numerous patient and acquisition-specific variables, such as age, sex, scanner hardware, sedation protocols, and coexisting medical conditions, alter disease presentation across individuals~\cite{bera2023noise}\cite{chen2023fedsoup}\cite{chen2023federated}\cite{chen2023treasure}\cite{hu2023devil}\cite{parker2023gender}. These sources of variation, collectively contributing to intraclass diversity~\cite{che2023towards}\cite{liu2023development}, pose difficulties for deep learning (DL) systems, particularly when only limited examples of certain pathologies are available, leading to reduced sensitivity to acute abnormalities~\cite{kim2023dimix}. Such performance degradation frequently stems from dataset imbalance when models are often trained on overwhelmingly non-pathological cases, causing the rare but clinically essential pathological samples to be underrepresented. Although these rare instances contain the most informative signals, their scarcity limits the model’s ability to learn robust representations. Consequently, identifying rare conditions necessitates specialized techniques that explicitly address their low prevalence, distinct visual markers, and nuanced patterns of positive and negative evidence, since purely data-driven statistical learning cannot fully capture the underlying distribution of rare classes from sparse observations~\cite{abubakar2024systematic}. Rare class detection is especially important in scenarios like detection of stage 5 diabetes retinopathy (DR) or identifying the seizure onset zone (SOZ) from resting state (rs) fMRI~\cite{che2023towards,kamboj2025generating}, or detecting coronary artery disease (CAD) from stress ECG and it remains a challenging task.

Researchers have proposed several strategies to mitigate domain generalization issues, including domain adaptation (DA) and multi-source domain generalization (MSDG), where training incorporates data originating from multiple external centers or source domains~\cite{hu2023devil}. Such approaches have demonstrated improved robustness when deep learning (DL) models are evaluated on datasets drawn from disparate clinical environments. Nevertheless, DA and MSDG frameworks typically require substantial effort to obtain data from those additional domains and often raise privacy concerns because they involve sharing or aggregating data across institutions and limit their real-world applicability in healthcare settings~\cite{hu2023devil}. Given these constraints, single-domain generalization (SDG) has emerged as a more practical alternative, focusing on achieving cross-domain robustness using only a single-source dataset~\cite{yan2023epvt,vidit2023clip}. Although SDG has been explored extensively for common image classification tasks, to the best of our knowledge, we find no existing work that tackles SDG specifically for rare-class scenarios, where the generalization problem is fundamentally more challenging.

Strategies enhancing SDG for rare categories can benefit from incorporating expert-derived insights, such as clinically recognized, domain-invariant cues that characterize disease signatures and help disentangle visually similar classes. These expert perspectives can also offer guidance on handling intra-class variability. Yet, in clinical practice, such knowledge is often imprecise, subjective, and influenced by individual interpretation~\cite{boerwinkle2017correlating, hunyadi2015prospective}. Moreover, medical datasets frequently contain measurement noise and acquisition artifacts, which may cause substantial overlap between disease presentations, further complicating the learning problem~\cite{kamboj2023expert,boerwinkle2017correlating}. As a result, many traditional knowledge-driven systems designed for healthcare applications exhibit considerable rates of false positives (FPs) and false negatives (FNs)~\cite{KambojTAI, kamboj2023expert, banerjee2022automated, nandakumar2023deepez, galappaththige2024generalizing}.

In this paper, we first share our motivating experience in deploying an AI technique for detection of CAD from stress ECG at Mayo Clinic that failed the SDG performance evaluation test~\cite{banerjee2025enhancement}. We then introduce a fundamentally different approach to SDG for rare classes. We utilize non-data driven, discriminative expert knowledge and integrate it with DL using pre-trained large vision model (LVM) and demonstrate superior SDG performance in rare class detection than SOTA on the commonly used SDG benchmark of diabetes retinopathy (DR). We then demonstrate the on-field performance of human expert integrated DL (DL+EKE) on two deplotment experiments on rs-fMRI based SOZ detection and CAD detection using stress ECG.  


Our main contributions are: i) We introduce integration of domain invariant expert knowledge with DL as a viable solution for rare classes detection and SDG, using LVM and class-wise entropy, overcoming the challenge of limited and imbalanced data, as well as overlapping information gathered from expert knowledge about classes. ii) Demonstrate superior performance of DL+EKE on DR benchmarks , and iii) extensive deployment experiments on two example medical imaging applications (SOZ and CAD detection) and comparisons with SOTA show that DL models, when augmented with expert knowledge and designed to break information overlap, effectively detect rare classes from disparate sources, thereby demonstrating enhanced generalizability using single domain.

\section{Related Work and Preliminaries}

\textbf{Domain Adaptation and Domain Generalization}.  
Domain adaptation (DA) aligns a source distribution to a known target, while multi‐source domain generalization (MSDG) uses multiple labeled sources to generalize to unseen targets. Both require either target or diverse source data, which is typically unavailable in medical settings~\cite{hu2023devil}. Single‐source domain generalization (SDG) learns invariant features from one labeled domain~\cite{vidit2023clip}, but performs poorly under class imbalance and limited rare‐class samples~\cite{li2023frequency,chen2023federated,kim2023dimix,yang2023learning}. Image‐ and feature‐level augmentations likewise underperform, as they sample from the same source distribution and fail to capture rare‐class variability.

\textbf{Knowledge and DL}.  
Incorporating expert knowledge into DL has involved feature encoding~\cite{banerjee2022automated} or domain‐specific augmentations~\cite{hu2023devil}, but overlapping clinical concepts (e.g., similar power‐spectra thresholds) limit discrimination~\cite{boerwinkle2017correlating}. Physics‐guided networks embed equations into the loss, yet falter when knowledge is vague. Knowledge‐enhanced neural networks impose symbolic output constraints~\cite{KENN}, but cannot model intra‐class variability and struggle with incomplete or inconsistent knowledge. Our method integrates knowledge with single‐domain data to address rare‐class imbalance and overlap without requiring target data.

\noindent{\bf Rare Class Properties} ``Rare classes are extremely infrequent classes whose characteristics make them or their consequences highly valuable. Such classes appear with extreme scarcity and are hard to predict, although they are expected eventually''~\cite{sokolova2010evaluation}. In this paper, we consider \textit{rare class} to be a phenomenon which has four properties: a) \textbf{Discrimination}: observations of the phenomenon have distinctive characteristics than other observations, b) \textbf{Scarcity}: the phenomenon has less number of observations in the process, c) \textbf{Significance}: each observation of the phenomenon has much more information content than other observations of the process, d) \textbf{Overlap}: each observation of the phenomenon has several characteristic features that exhibit significant overlap or high similarity within the feature embedding space, independent of class labels. This means that the embedding space is not exposed to the specific classes of the domain in question.

\section{Motivating Deployment Experience}
The motivating problem was the development of an AI technique to detect CAD from a patient’s 12-lead exercise stress ECG (across different exercise stages). The ground truth for the presence of obstructive CAD ($\geq$ 70\% stenosis), is obtained using invasive coronary angiography (ICA) performed within ~2 months of stress ECG. A visual transformer (ViT, discussed in Appendix)  based architecture was trained with data from the year 2010 on 1200 patients (1000 train and 200 validation) with equal distribution of CAD positive and negative ICA outcomes. The ViT was then tested on an untouched data for 200 patients from the 2010 repository. The positive or negative predictive value (PPV or NPV) of this ViT model as shown in Table \ref{tbl:ViT} are excellent on the test data with very little deviation from validation. However, when the ViT was tested on 92 patients from 2025, the PPV and NPV dropped drastically (Table \ref{tbl:ViT}). 

\begin{table}[htbp]
\centering
\scriptsize
\label{tab:knowledge_ablation}
\begin{tabular}{@{}p{0.3 in}@{}p{0.4 in}@{}p{0.8 in}@{}p{0.6 in}@{}p{0.8 in}@{}p{0.4 in}@{}}
\hline
\textbf{Metric} & \textbf{Method} & \textbf{Validation (y2010)} & \textbf{Test (y2010)} & \textbf{Blind Test (y2025)} &\textbf{Reduction} \\
\hline
PPV & ViT & 80.4\% & 79.0\% & 46.0\% & 33\% \\
NPV & ViT & 83.0\% & 81.8\% & 49.0\% & 32.8\%\\
\hline
\end{tabular}
\caption{ViT performance under \emph{single-domain} training: 1{,}000 patients; Validation: 200; Test: 200; Blind Test: 92.}
\label{tbl:ViT}
\end{table}

Prior to 2012, if
clinicians did not find S-T depression as a feature in the stress ECG image, then patients were not
referred to ICA. However, stress ECG
has other subtle positive CAD features manifested in the inter-relationship among time series from
different leads which do not cause any data distribution shift in individual leads and can be easily
ignored by manual labeling. A change in triage policy in 2012 resulted in more patients being sent for
ICA even if S-T depression was not found. Hence, in data from 2025, there were CAD positive stress ECG
that had expert knowledge factors (inter lead relationship) affecting CAD diagnosis that were not present in data
from 2010.

\begin{figure*}[h]
\centering
\includegraphics[width=0.85\textwidth]{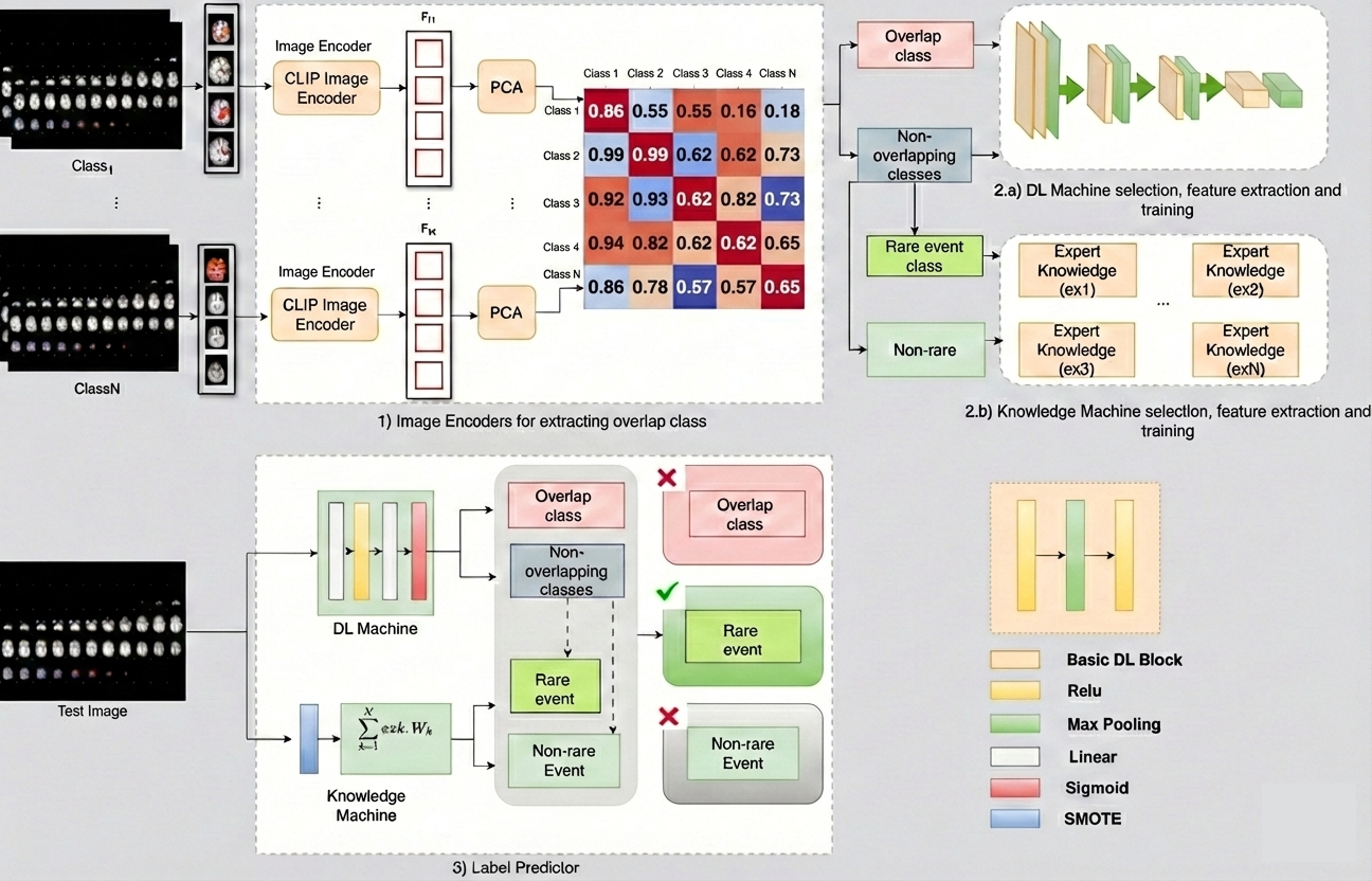}
\caption{\textbf{RareSaGe Model Schematic:} \textbf{1) Image Encoders:} CLIP identifies the class most similar to the rare class. \textbf{2.a) DL Machine} distinguishes overlap and non-overlap classes. \textbf{2.b) Knowledge Machine} extracts features from the rare class, and trains a quadratic optimization model. \textbf{3) Label Predictor:} Each test image is first classified by the DL classifier as either overlap or non-overlap. Non-overlapping images are further classified by the knowledge machine as rare or non-rare.} \label{fig1}
\end{figure*}

\section{Methods}
\label{section3}


Our approach to solving the SDG rare class problem in Definition \ref{dfn:Prob1} involves integrating expert knowledge with DL techniques. First, we consider the source domain $Y_1$ and isolate the rare class from the other available classes in the raw data, following the criteria outlined in Definition~\ref{def:Rare}. For this purpose, we need a representation $x_i$ of the observations $y_i \in Y_1$. For DL, we employ large vision model (LVM) CLIP to extract features from the raw data, independent of class information (Fig.~\ref{fig1})~\cite{radford2021learning}. For expert knowledge, we utilize symbolic AI based representations highlighted in Section "Expert Knowledge on rs-fMRI". This step is application dependent and requires specialized ML methods. We compute the similarity between the class agnostic feature embeddings of the rare class and each of the classes in  $C$, referring the most similar class to the rare class as the overlap class $c_o = argmin(\theta_i)$. We use DL techniques to classify overlap class, and knowledge for rare class. We utilize a machine orchestration strategy to derive the best DL and knowledge based machine. This overall strategy is formulated in algorithm RareSaGe.


\noindent{\bf Machine orchestration strategy:} We assume that there is a set of trained DL/ML classifiers $\mathcal{M_{DL}}$ for the overlap and non-overlap classes' learning, and a set of trained knowledge based classifiers $\mathcal{M_{K}}$ for the knowledge-based rare class and non-rare classes (normal classes) learning. Each classifier $M_d \in \mathcal{M_{DL}}$ or $M_d \in \mathcal{M_{K}}$, classifies instances of $Y$ with a label from the set $S^{M_d} \subset 2^C$, such that each label in $S^{M_d}$ meets the following rules: \textbf{a) Mutual exclusion:} {no two labels in $S^{M_d}$ share instance from the same original classes $c_i \in C$}; \textbf{b) Class cover:} the union of class labels $S^{M_d}$ includes all original classes; \textbf{c) Union rule:} the classes im $S^{M_d}$ can be expressed as union of classes from $C$.

  A classifier $M_d$ represents the raw data $y_i \in Y$ in some latent representation space $x_i \in X$ using a discriminative feature function $\mathcal{F}_{M_d}$. The same function can be used to represent each raw data in the original class set $C$. We select the machine that reduces rare class entropy (concept described in our prior work~\citep{KambojTAI}) the maximum amount. Following the entropy calculation in Eqn. \ref{eqn:theta} we can derive the entropy $\theta^{M_d}_r$ for rare class $c_r$ in classifier $M_d$ and select the classifier with the least entropy.

\noindent{\bf RARE class classification for Single domain Generalization (RareSaGe) overview:} Algorithm \ref{alg:eksaii} in Appendix proposed in~\cite{KambojTAI,kamboj24Storm} is used to derive a classifier for the rare class $c_r$. 







\noindent{\bf Multiple Rare Classes:} The Algorithm \ref{alg:eksaii} can be applied iteratively on multiple rare classes one by one. We show this with three examples: single rare class CAD and SOZ detection, and multiple rare class DR grading.


\subsection{Application of RareSaGe}
We show RareSaGe application to SOZ detection problem. Stress ECG and DR grading examples are in appendix.






In standard pre-surgical screening, resting-state fMRI (rs-fMRI) is acquired, producing 4D spatio-temporal data. Independent component analysis (ICA) decomposes rs-fMRI into three classes of spatial–temporal independent components (ICs): (i) noise ICs capturing motion and measurement artifacts, (ii) resting-state network (RSN) ICs representing normal brain activity, and (iii) seizure onset zone (SOZ) ICs corresponding to epileptogenic activity. Each fMRI scan typically yields 100–200 ICs per patient, of which only a small subset ($\approx 5-10\%$) correspond to SOZ ICs, constituting a rare-class detection problem~\cite{kamboj2023expert,banerjee2022automated}. Fig. \ref{fig1} shows the architecture of expert knowledge integration with DL for rare class detection. 

\subsubsection{Deep Learning for noise}

DL techniques, including Vision transformers (ViT), language vision models (LVM)s~\cite{radford2021learning}, 2D-CNNs, and transfer learning using pre-trained models are effective at capturing intricate spatial patterns and features within images~\cite{he2016deep}. Given the prevalence of noise class instances in medical imaging datasets, often constituting over 50\% of the data~\cite{boerwinkle2017correlating, banerjee2022automated, kamboj2023expert}, DL can leverage its capability to detect subtle spatial patterns and features that distinguish noise from other classes. 

\subsubsection{Expert Knowledge on rs-fMRI} 
\label{sec:Expert}

We utilize two types of expert knowledge on rs-fMRI data: 

\noindent{\bf a) Anatomical knowledge:} This pertains to the spatial locations of anatomical brain parts crucial for SOZ recognition. These locations can be extracted employing established image processing algorithms. Locations are in Appendix.

\noindent{\bf b) Expert knowledge on specific rare class:}  This encompasses knowledge about rs-fMRI activation patterns observed for SOZ, compiled from the works of Hunyadi et al.~\cite{hunyadi2015prospective} and Boerwinkle et al.~\cite{boerwinkle2017correlating}. SOZ specific knowledge is expressed as logical connectives of atomic propositions on the location of activation relative to brain anatomical regions (Appendix). The atomic proposition valuations are used in a support vector machine (SVM) based expert knowledge extractor (EKE).  

\subsubsection{Integration of expert knowledge with DL using Algorithm \ref{alg:eksaii}}
Following Algorithm \ref{alg:eksaii} we first identify rare class. From domain $Y_A$, the cross entropy using CLIP features for Noise was $\theta^{CLIP}_{Noise} = 0.004$, for RSN was $\theta^{CLIP}_{RSN} = 0.0046$, and for SOZ it is $\theta^{CLIP}_{SOZ} = 0.026$. We identify that SOZ is the rare class since $\theta^{CLIP}_{SOZ}$ satisfies Definition \ref{def:Rare}. Further, the cosine similarity of CLIP features between Noise and SOZ was $0.78$ while that between RSN and SOZ was $0.74$. Hence, the Noise class is determined to be the overlap class. We divide the dataset into $NOISE$ and $\neg NOISE$ class. Since we do not have knowledge based machines for Noise, we utilize DL techniques to classify $NOISE$ and $\neg NOISE$. The $\neg NOISE$ class is then used to determine SOZ. Here we have SOZ specific discriminative expert knowledge and we use the knowledge based machines to identify SOZ. We integrate expert knowledge with DL in the post processing step (Fig \ref{fig1}, Label Predictor). The DL first classifies IC as Noise (overlap) or non-Noise (non-overlap) class. Simultaneously, the EKE classifies IC as SOZ (rare) and RSN (non-rare/normal). If DL classifies an IC as noise and EKE categorizes it as SOZ with a confidence score surpassing a selected threshold of 0.9, the IC is labeled as SOZ; otherwise, it retains noise label. For ICs labeled as non-noise by DL, the EKE label of RSN or SOZ is selected as the final label.

\begin{table*}
\scriptsize
\centering
\begin{tabular}{lllllllll|} 
\hline
\multicolumn{2}{|l|}{\textbf{Method}} & \textbf{Accuracy} & \textbf{Precision} & \textbf{Sensitivity} & \textbf{F1-score} & \textbf{Average F1-score} & \textbf{Time(min)} & \textbf{Ablation}\\
\hline

\multicolumn{2}{|l|}{Pre-trained ViT small Train A, Test B}       & 64.5\% & 86.9\% & 71.4\% & 78.4\% & 77.2\% & 45(12) & DL only \\
\multicolumn{2}{|l|}{Pre-trained ViT small Train B, Test A}       & 61.5\% & 91.4\% & 65.3\% & 76.1\% &         & 42(13) & DL only \\ \hline





\multicolumn{2}{|l|}{Knowledge based system, Train A, Test B}     & 83.8\% & 89.6\% & 92.8\% & 91.2\% & 78.9\% & 13(6)  & Knowledge only \\
\multicolumn{2}{|l|}{Knowledge based system, Train B, Test A}     & 50.0\% & 89.6\% & 53.0\% & 66.6\% &         & 12(6)  & Knowledge only \\ \hline

\multicolumn{2}{|l|}{\textbf{DL+EKE Train A, Test B}}             & \textbf{90.3\%} & \textbf{90.3\%} & \textbf{100\%} & \textbf{94.9\%} & \textbf{90.2\%} & \textbf{35(10)} & \textbf{DL+EKE} \\
\multicolumn{2}{|l|}{\textbf{DL+EKE Train B, Test A}}             & \textbf{75.0\%} & \textbf{92.8\%} & \textbf{79.5\%} & \textbf{85.6\%} &         & \textbf{38(9)}  & \textbf{DL+EKE} \\ \hline
\end{tabular}
\caption{Performance results of across-trial validation—single domain generalizability for rare class detection. Here we show only the best DL only approach. For all other baselines refer to Table \ref{tab2} in Appendix}
\label{tab1}
\end{table*}


\section{Experiments and Results}
\noindent{\bf DR grading:} benchmarks and baselines are in Appendix.
\subsection{Deployment description}
\noindent{\bf Deployment 1, SOZ detection:} DL+EKE was trained using data from, Phoenix Children's Hospital (Center A) and deployed for testing in University of North Carolina (Center B), in compliance with IRB protocols and cross-university agreements. Center A includes 52 pediatric patients (23 Male, 29 Female, ages 3 months to 18 years) with 5,616 images (2,873 Noise, 2,427 RSN, 316 SOZ), acquired using a 3T Philips Ingenuity scanner. Test Center B includes 31 patients (14 Male, 17 Female, ages 2 months to 62 years) with 2,364 images (1,090 Noise, 1,072 RSN, 202 SOZ), acquired using a Siemens MAGNETOM Prisma FIT scanner. 

\noindent{\bf SOZ baselines:} There is no existing DL based baseline available for SOZ detection. We develop our own DL techniques for comparison including: pre-trained CNN using “VGG-16,” pre-trained ViTs using “vit-small-patch16-224” and Google's “vit-base-patch16-224,” ViTs trained from scratch with hyperparameters optimized using Optuna, LVM CLIP~\cite{radford2021learning}, and knowledge-based systems.

\noindent{\bf Deployment 2, CAD detection:} The Collaborative Institutional Review Board (IRB) provides access to the Mayo Integrated Stress Center (MISC) database, containing over 100,000 Exercise Stress ECG (ESE) cases linked to invasive coronary angiography (ICA) from 2010. The dataset covers CAD patients from three centers~\cite{banerjee2025enhancement}.

\noindent{\bf Blind testing:} Subsequently, the model was re-tested on a second validation cohort (n = 92) through blind testing where the labels of the stress ECG were not disclosed to the developers at ASU during inference. The labels were manually verified by the collaborators at Mayo Clinic. The distribution of CAD in this cohort was not predefined by the investigators. For this second cohort, random patients who underwent exercise stress ECG and coronary angiography within 3 months between February 2025 and May 2025 were included. Patients previously included in the first cohort, those with overlapping ECG waves or missing data, and patients who experienced acute coronary syndromes between the stress ECG and angiogram were excluded

\subsection{Experiments} \label{ID}
To evaluate generalizability, we perform two experiments motivated from~\cite{chekroud2024illusory}: 


\noindent{\bf A) Across trial validation:} Here, $M_{DL(N)}$ and $M_{EKE(N)}$ is trained on data of center $X$ with $N$ patients and tested on dataset of the other centers $Y\neq X$ with $P$ patients. This trial tests the SDG performance. 

\noindent{\bf B) Aggregated trial validation:} This is leave-one-domain-out method where data from all centers but one combined ($A \bigcup B$) was evaluated using 5-fold, 3-repeats validation. 
\subsection{Evaluation metrics, Results and Insight}
\noindent{\bf Evaluation metrics:} Trials are evaluated with standard metrics used in~\cite{hunyadi2015prospective, KambojTAI} such as average accuracy of multi-class classification and F1 score for identifying rare class. For the SOZ detection example, we evaluate average execution time in detecting SOZ from each patient. This quantifies computational complexity of our technique and we compare it with SOTA. For CAD detection we present positive/ negative predictive value (PPV/NPV) which are clinically relevant. 

\noindent{\bf DR grading results:} Tables~\ref{tab:eyepacs_source} through~\ref{tab:messidor2_source} in Appendix present systematic evaluation across all 12 source-target pairs for across trial validation SDG performance. DL+EKE consistently outperforms SOTA benchmarks. Table~\ref{tab:vlm_performance} in appendix presents fair comparisons where all methods are trained on three datasets and evaluated on the fourth for aggregate trial validation. DL+EKE demonstrates substantial improvements over SOTA domain generalization approaches. DL+EKE's superior performance stems from the integration of medical vision-language pre-training with domain conformal boundaries. 


\noindent{\bf SOZ detection results:} Table \ref{tab1} presents the results of state-of-the-art DL techniques and our methodology on the across-trial validation test, as discussed in Section \ref{ID}. All pre-trained models were fine-tuned on data from specific centers, with loss updated using class weights to address dataset imbalances due to rare classes, and early stopping strategies applied to prevent overfitting. For LVM, we conducted two evaluations: one fine-tuning with contrastive loss and another with cross-entropy loss, both using the “ViT-B/32” model. Our methodology utilized a 2D-CNN as a DL machine, which performed best for evaluating overlap (noise) and non-overlap (non-noise) classes (Fig.~\ref{fig1}). In the across-trial experiment, our DL+EKE method achieved an F1 score of 90.2\%, consistently maintaining F1 scores above 85.0\% for both centers. This significantly outperforms other baseline comparators, highlighting our technique’s generalizability to unseen data and its ability to learn domain-invariant expert knowledge without overfitting. Additionally, in the aggregated trial validation, where models were exposed to more data from both centers, Table~\ref{table2} in Appendix demonstrates that our technique achieves the best results, with an average F1 score improvement of 4.5\%.  

\noindent{\bf SOZ detection insights:}
Our across-trial experiments reveal that state-of-the-art DL techniques struggled to differentiate between NOISE and SOZ classes, as indicated by CLIP similarity results, leading to suboptimal performance in rare class detection. A knowledge-based system trained on Center A accurately identified SOZ characteristics in Center B with high precision. However, when trained on Center B and tested on Center A, precision remained stable, but sensitivity dropped significantly, highlighting an increase in FNs due to patient variability. Integrating DL for overlap class separation with knowledge-based methods improved both FPs and FNs, resulting in a more generalized performance. The aggregate trial achieved higher F1 score of 94.7\%.

\begin{figure}
\centering
\includegraphics[width=0.4\textwidth]{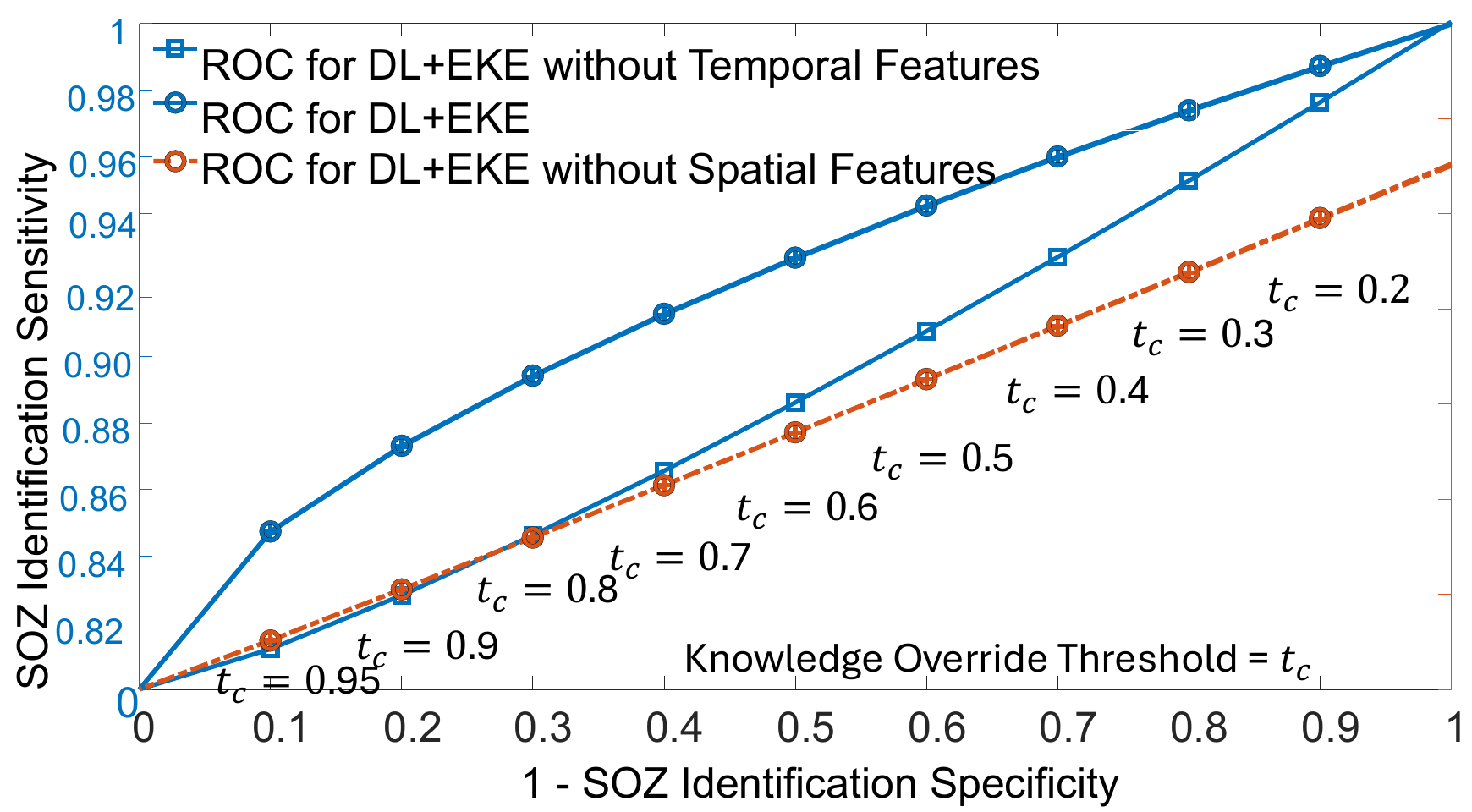}
\caption{ROC of DL+EKE for different knowledge override thresholds and ablation analysis of knowledge.} 
\label{sense}
\end{figure}

\noindent{\bf Ablation studies:} Table \ref{tab1} shows that DL+EKE performs better than DL only and knowledge only approaches. Fig. \ref{sense} further explores ablation of knowledge types. It shows that temporal knowledge components are not as important as spatial components since removing them makes the DL+EKE technique as good as random choice. 

\noindent{\bf Sensitivity to knowledge override threshold $t_c$:} The receiver operating characteristics (ROC) curve shows the variance of sensitivity and specificity of DL+EKE as the knowledge override threshold is varied from $0.1$ to $0.95$. It shows determining this threshold is not straightforward and reduction of this threshold may hamper performance.





\noindent{\bf CAD detection results:} We test four different configurations of DL+EKE, the best performing K1 and subsequent ablations: \textit{a) 5-Lead Transformer at Max METs (K1 - comparator):}
This primary architecture integrates the highest expert knowledge level, utilizing data from all maximum MET levels and restricting the lead count(L=5). \textit{b) 5-Lead Transformer Across All METs (K2):}
Maintaining a 5-lead constraint while including data from all MET levels. \textit{c) 12-Lead Transformer at Max METs (K3):}
Expanding the lead coverage to all 12 leads (L=12) focuses solely on data from maximum MET levels. \textit{d) 12-Lead Transformer Across All METs (K4):}
It uses all leads and all speeds.

\begin{table}[htbp]
\centering
\scriptsize
\begin{tabular}{|c|c|c|c|c@{}|}
\hline
\textbf{Metrics} & \textbf{Method} & \textbf{Validation} & \textbf{Test Data (2010)} & \textbf{Blind Test (2025)} \\ \hline

\multirow{7}{*}{PPV}
& ViT     & 80.4\%               & 79\%               & 46.0\%  \\ \cline{2-5}
& K1     & 91.2\% (3.3)         & 91.2\%             & 75.0\%  \\ \cline{2-5}
& K2      & 87.2\% (3.0)         & 88\% (0.04)        & 74.0\%  \\ \cline{2-5}
& K3      & 83.3\% (4.3)         & 82.4\% (0.001)     & 72.0\%  \\ \cline{2-5}
& K4      & 83.4\% (3.0)         & 82.3\% (0.003)     & 71.5\%  \\ \hline

\multirow{7}{*}{NPV}
& ViT     & 83\%                 & 81.8\%              & 49.0\%  \\ \cline{2-5}
& K1     & 93.0\% (1.5)         & 93\%                & 76.0\%  \\ \cline{2-5}
& K2      & 85\% (22.3)          & 87\% (0.07)         & 73.0\%  \\ \cline{2-5}
& K3      & 90\% (22.4)          & 81\% (0.045)        & 74.0\%  \\ \cline{2-5}
& K4      & 83\% (17.2)          & 79\% (0.006)        & 72.0\%  \\ \hline
\end{tabular}
\caption{Comparison with SOTA unguided-ViT models, and DL+EKE models on validation and test performance.}
\label{tab:knowledge_ablation}
\end{table}

\noindent{\bf CAD detection insights:} Among the four expert-guided configurations, K1, incorporating five expert-selected leads and maximum METs, achieved the best performance. Relative to K1, using all METs reduced PPV by 4\% and NPV by 8\%, while using all leads with maximum METs reduced PPV by 7.9\% and NPV by 3\%. Using all leads and all METs further reduced PPV by 7.8\% and NPV by 10\%. For unseen test data, K1 achieved a PPV of 91.2\% and an NPV of 93\% (Table.~\ref{tab:knowledge_ablation}), representing a substantial improvement over the state-of-the-art PPV of 77\%. Additionally, expert-guided AI outperformed unguided approaches, yielding average gains of 12.2\% in PPV and 11.2\% in NPV. The 5-fold cross-validation ROC AUC was 92.2 ($\pm$ 1.1), and knowledge guided methods demonstrated strong SDG performance.

\begin{figure}
  \centering
\includegraphics[trim=0 100 0 0,width=0.45\textwidth]{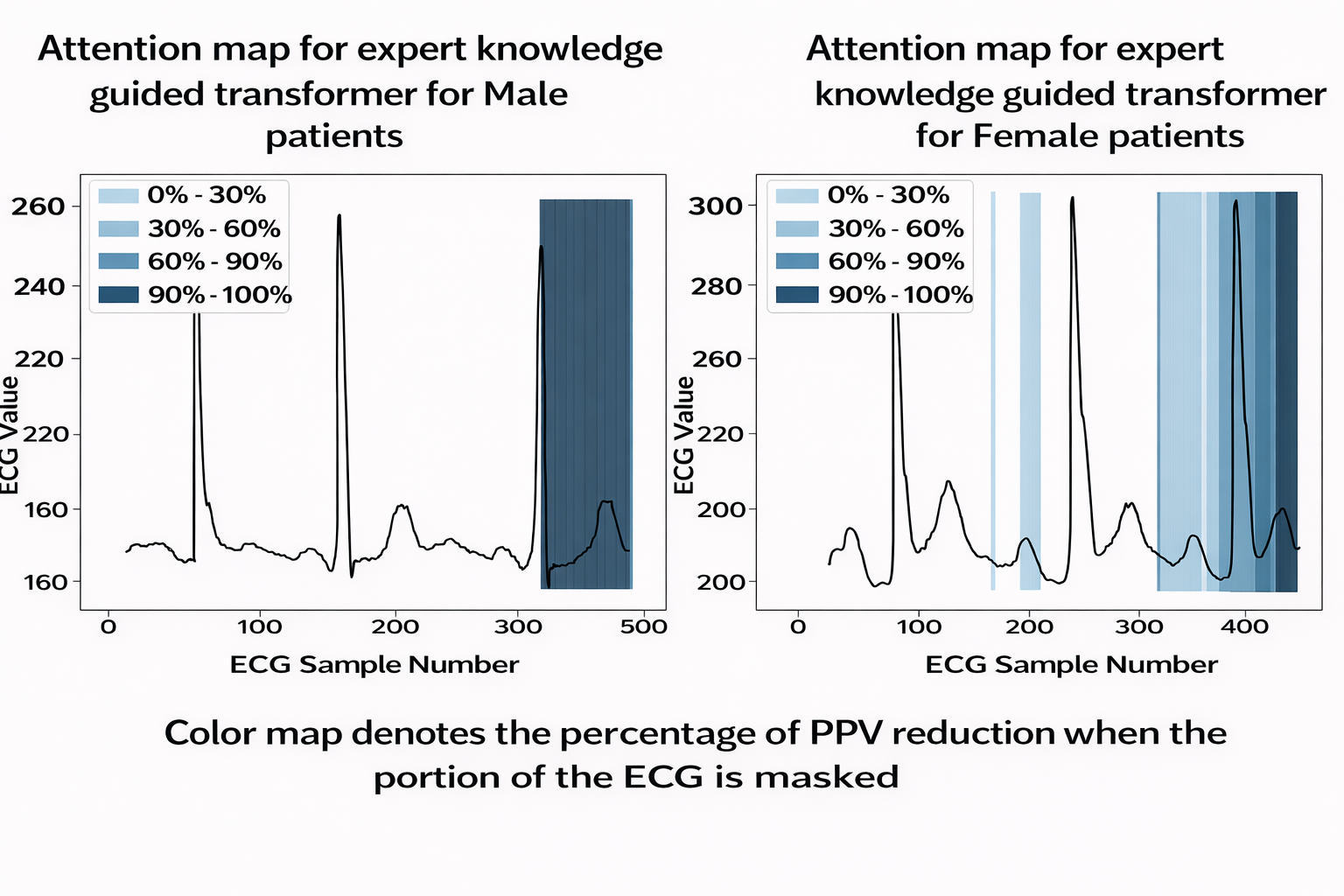}
  \caption{LIME attention maps of DL+EKE focuses on ST segment for males, but on QRS and P waves for females.} 
  \label{fig:Fig4}
\end{figure}

\section{Discussion on Deployment Experience}
The \textbf{main} observation is that state-of-the-art single-domain generalization methods fail to generalize across both deployment applications and the DR benchmark, and \textbf{do not achieve clinically significant performance}.

Incorporating expert knowledge can improve generalization, but comes at the \textbf{cost of substantial manual effort} to identify and encode domain expertise. Moreover, many inter-center variations are difficult to anticipate without a collaborative development environment that enables continuous interaction between clinical and engineering teams. The absence of such interaction can significantly delay deployment-ready AI. Clinicians often require clinically meaningful parameters that are not exposed by standard AI/ML pipelines; retrofitting explainability after development can necessitate extensive re-engineering and must therefore be addressed early in the design phase.

\noindent{\bf Post deployment interaction at Mayo Clinic:} After the blind testing on 92 patients, the research team presented the technique and results to team of experts in internal medicine and cardiology. Interaction with experts resulted in the following questions and suggestions: 

\noindent{\bf Q1. Can the attention map explain why a positive manual stress ECG interpretation was a false positive (not confirmed in ICA)? } DL+EKE has the capacity to show the part of a signal that has a domnant effect on the outcome. However, this component is not thoroughly validated. 

\noindent{\bf Q2. Can DL+EKE grade data quality and qualify the recognition result with a confidence?} DL+EKE assumes good quality stress ECG images where the leads are not overlapped. The performance of DL+EKE with noisy stress ECG is yet to be quantified. 

\noindent{\bf Q3. Can the attention map educate a trainee about signs of a positive stress ECG result that is confirmed by ICA?} The question indicates an interest in using DL+EKE as an AI tutor for ECG readers. While this is theoretically possible, DL+EKE was not originally designed for this purpose.

\noindent{\bf What signal characteristics did DL+EKE use to determine CAD positive results for women?} LIME~\cite{simsar2024limelocalizedimageediting} based attention maps of the transformer is shown in Fig. \ref{fig:Fig4}. The question shows strong interest in reducing the disparity between men and women. This delves into ethical use of AI where AI can potentially reduce health disparities.

\section{Conclusions}
Single domain generalization is an essential property of deployment ready AI. SOTA domain generalization techniques may show improvements in benchmarks but are nowhere near clinically relevant performance that can be readily deployed in patient facing applications. Expert knowledge can improve generalized performance however, knowledge is vague and often conflicting between experts. Hence, knowledge refinement is essential which can only occur through iterative dialogue early in the development process. There needs to be a strong theoretical exploration on the reasons for failure of SOTA domain generalization methods. A formalism of the core reason of causal changes across domains may result in more efficient and better SDG methods. This is essential for deployable AI in the medical field.



 \bibliography{miccai}
 \newpage
 \clearpage
\appendix

\section{Appendix}
This section gives the details that could not be included in the main paper for space limitations.

\section{A1: Definitions}

Observations of the process can be categorized into a finite number $(m)$ of classes that are common across domains. A \textit{rare class}, $c_r \in \{c_1,\ldots,c_m\}$ is defined as a collection of \textit{rare artifacts} across all domains.

\begin{definition}[SDG problem definition] \label{dfn:Prob1}
Consider at least two mutually exclusive datasets $Y_1$ and $Y_2$ ($Y_1 \bigcap Y_2 = null$) from two different domains, such that each instance from each domain belong to a unique class out of the set of $m$ classes $C=\{c_1 \ldots c_m\}$. Using only observations from dataset $Y_1$, determine the label of test data $t_y \in Y_2$.
\end{definition}

\noindent{\bf Rarity Quantification:} We quantify rare class using class wise entropy metric \cite{b59}. We define distance $dist(x_i, x_j)$, between representations $x_i$ and $x_j$ of two observations $y_i$ and $y_j$ in dataset $Y$ with $n$ observations in a given domain $z \in \{1 \ldots q\}$ ($q \geq 2$ domains) with distribution function $\mathcal{D}_z(c_k,\omega_k)$ for any class $c_k$ with class specific parameter set $\omega_k$. For each observation $y_i$ of a rare class $c_r$, we define $Q(x_i)$ as the set of all observations $y_j$ such that $x_j, x_i \in c_r$, and $y_j$ is in the $K$ nearest neighbor set of $y_i$ using the distance metric $dist(x_i,x_j)$. The set $Q(x_i)$ defines density of $y_i$, $\lambda(x_i)$.

\begin{equation}
\scriptsize
\label{eqn:dist}
\lambda(x_i) = \frac{1}{|Q(x_i)|} \sum_{j=1}^{|Q(x_i)|} \frac{1}{\text{dist}(x_i, x_j)},
\end{equation}
where \(|Q(x_i)|\) is the number of observations in the set \(Q(x_i)\).  The class average density of an observation \(y_i\) $\in$ \(c_r\) is,

\begin{equation}
\scriptsize
\label{eqn:dist2}
\gamma(x_i) = \frac{\lambda(x_i)}{\sum_{j=1}^{|c_r|} \lambda(x_j)},
\end{equation}
The class entropy for \(c_r\) is defined using Eqn. \ref{eqn:theta},
\begin{equation}
\scriptsize
\label{eqn:theta}
\theta_r = \sum_{i=1}^{|c_r|} (-\gamma(x_i) \log_2 \gamma(x_i)).
\end{equation}

\begin{definition} \label{def:Rare}
In this paper, we define a class $c_r$ as a rare class if, for all domains $z \in \{1 \ldots q\}$, $\theta^z_r$ ($\theta_r$ for a domain $z$) falls outside $2\sigma^z_\theta$ range from the mean value $\theta^z_M$ of $\theta$ across all classes in domain $z$, where $\sigma^z_\theta$ is the standard deviation of the class entropy values across all classes, i.e. \\ $\theta^z_r > \theta^z_M + \sigma^z_\theta \; or \; \theta^z_r < \theta^z_M - \sigma^z_\theta$, \\ where, $\theta^z_M = \frac{\sum_{i = 1}^{m}{\theta^z_i}}{m}$ and $\sigma^z_\theta = \sqrt{\frac{\sum_{i = 1}^{m}{(\theta^z_i - \theta^z_M)^2}}{m}}$.
\end{definition}

\noindent{\bf Rationale for Definition \ref{def:Rare}:} The fundamental reason machine learning (ML) techniques may fail in solving Problem Definition \ref{dfn:Prob1} is that there are not enough information available in the rare class observations to estimate the true distribution $\mathcal{D}_z(c_r,\omega_r)$, where $\omega_r$ is the parameter of the distribution for class $c_r$~\cite{abubakar2024systematic}. The lower bound of the error in estimating the parameters of a distribution using an unbiased estimator is given by the Cramer Rao Lower Bound (CRLB) as: 
\begin{equation}
\label{eqn:CRLB}
\scriptsize
    e_{\omega_r} \geq 1/I(\omega_r),
\end{equation}
where $I(\omega_r)$ is the Fischer information~\cite{lin2005cramer}. Barron et al.~\cite{Barron86Entropy} provides a close form relationship between class entropy $\theta$ computed using Eqn. \ref{eqn:theta}  and Fischer information $I(\omega)$, where as $\theta$ increases, $I(\omega)$ should decrease. According to the CRLB theory, this implies that as $\theta$ increases $e_\omega$ increases. Hence, for the rare class, since class entropy is higher, the error in estimating the distribution by an unbiased estimator is higher, i.e., $e_{\omega_r} > e_{\omega_i}$, $\forall i \in \{1\ldots m\}, i\neq r$. 

The above-mentioned result implies that a direct ML/DL based approach towards solving SDG for rare class problem in Definition~\ref{dfn:Prob1} may not be successful, since it will have higher error in detecting rare class than other classes. 

\section{A2: RareSaGe Algorithm}
\begin{algorithm}
\caption{Conceptual RareSaGe Workflow in a clinical}
\label{alg:eksaii}
\scriptsize
\textbf{Input:} Patient data $Y$, list of conditions $\{c_0,\ldots,c_r,\ldots,c_n\}$, DL classifiers $\mathcal{M_{DL}}$, expert-rule classifiers $\mathcal{M_{K}}$, confidence threshold $t_c$.\\
\textbf{Output:} Predicted clinical condition

\begin{algorithmic}[1]

\STATE \noindent{\bf Step 1: Identify the Rare Condition}\\
Determine which condition $c_r$ is considered rare based on low frequency and high clinical importance.

\STATE \noindent{\bf Step 2: Find the Most Similar Common Condition}\\
Locate an “overlap” condition $c_o$—the common condition whose clinical appearance most resembles the rare one.

\STATE \noindent{\bf Step 3: Choose the Best Diagnostic Engine for Overlap Cases}\\
Compare deep-learning models and expert-rule models on distinguishing $c_o$ from “not-$c_o$” cases.\\
Select method with lowest entropy.

\STATE \noindent{\bf Step 4: Separate Out the Overlap Cases}\\
Use the selected method to classify and remove $c_o$ cases from the rest, leaving a pool containing $c_r$ and non-rare cases.

\STATE \noindent{\bf Step 5: Choose the Best Diagnostic Engine for the Rare Condition}\\
Again compare deep-learning and expert-rule models—this time for separating $c_r$ from “not-$c_r$” within the remaining data.\\
Select the method that performs most reliably for the rare condition.

\STATE \noindent{\bf Step 6: Combine Expert Knowledge and Deep Learning}\\
For final diagnosis:  
\begin{itemize}
    \item If the expert-rule system is confident ($> t_c$), trust the expert-rule decision.  
    \item Otherwise, rely on the deep-learning model’s prediction.
\end{itemize}

\STATE \noindent{\bf Output:} Final combined prediction for the patient’s condition.

\end{algorithmic}
\end{algorithm}

\section{A3: Expert knowledge on SOZ}

\noindent{\bf Anatomical Knowledge}

\noindent {\bf 1. Brain periphery:} This is the brain boundary obtained through Sobel filter based contour detection mechanism~\cite{banerjee2022automated}.\\
    \noindent {\bf 2. Gray matter:} Extraction is achieved through a combination of three methods: Global Probability of Boundary, Oriented Watershed Transform, and Ultrametric Contour Map as detailed in~\cite{banerjee2022automated}. This method is applied on the gray scale image to extract the gray colored areas of the brain.\\
    \noindent {\bf 3. White matter:} Similar extraction method as gray matter, focusing on white regions within the grayscale image. \\
    \noindent {\bf 4. Vascular regions:} Areas of blood vessels in the brain, extracted through slice processing techniques detailed in~\cite{banerjee2022automated}.\\
    \noindent {\bf 5. rs-fMRI activation:} Activation in the rs-fMRI image is obtained through DBSCAN, with minPoints parameter set to 2 voxels (where a voxel represents the smallest activation unit, typically a 3 pixel $\times$ 3 pixel image segment), and a neighborhood distance set to 1 voxel. All activation clusters below 135 voxels are disregarded as weak.

\noindent{\bf Expert knowledge on specific rare class for SOZ:}

    \noindent $p_1:$ Presence of a single activation cluster inside brain boundary, computed by counting the clusters of size greater than 135 voxels, completely inside the brain contour.\\
    \noindent $p_g:$ Activation primarily located in gray matter, calculated by the percentage of voxels within the activation lying inside the gray matter contour using the winding number algorithm~\cite{jacobson2013robust}.\\
    \noindent $p_s:$ Sparse representation of the blood oxygen level dependent (BOLD) signal in the sine domain, computed using Gini Index on the BOLD signal frequency response of each IC~\cite{hunyadi2015prospective}.\\
    \noindent $p_a:$ Activelet domain representation of the BOLD signal is sparse. This is also computed by first performing activelet transform and then computing sparsity using Gini Index~\cite{hunyadi2015prospective}.\\
    \noindent $p_w:$ Activation overlaps with white matter. This is computed using the same method as $p_g$, but on white matter contours.\\
    \noindent $p_v:$ Activation overlaps with vascular regions. This is computed using the same method as $p_g$, but on vascular contours.

Utilizing the aforementioned atomic propositions, the expert knowledge on SOZ rare classes can be encapsulated in the following first-order logical formula:
\begin{equation}
\scriptsize
    \label{eqn:refined}
    \kappa_{SOZ} = p_1 \wedge \neg p_s \wedge p_a \wedge [p_g \wedge (\neg p_w \vee (p_w \wedge p_v))],
\end{equation}
which represents the knowledge that SOZ activations primarily have one big activation cluster, the BOLD signal for SOZ ICs are not sparse in sine domain but are sparse in activelet domain, activation should primarily be in gray matter with no white matter overlap, or they may span across gray matter, white matter, and vascular regions.

\section{A4: ViT for stress ECG}

\label{sec:ecg_vit}

\subsection{Input Representation}
Let a patient record be a tensor
\[
\mathbf{X}\in\mathbb{R}^{S\times L\times T},
\]
where $S$ is the number of exercise stages (e.g., MET levels kept near peak), $L{=}12$ ECG leads, and $T$ is the per–stage duration in samples (resampled to a fixed horizon, e.g., $T{=}1000$–$2000$ at 500\,Hz). Preprocessing includes bandpass filtering (0.5–40\,Hz), baseline-wander removal, and per–lead $z$-normalization. Optionally, we retain the top-$K$ stages by workload (e.g., peak$\pm$1), but the formulation admits all $S$.

\subsection{Tokenization (Patchification)}
We treat each stage slice $\mathbf{X}_s\in\mathbb{R}^{L\times T}$ as a 2D ``lead$\times$time'' map and extract non-overlapping patches of size $p_h\times p_w$ (leads $\times$ samples). For stage $s$ this yields $N_s=\lceil L/p_h\rceil\cdot\lceil T/p_w\rceil$ patches $\{\mathbf{P}_{s,i}\}_{i=1}^{N_s}$, each flattened and linearly projected:
\[
\mathbf{t}_{s,i} \;=\; \mathbf{W}_p\,\text{vec}(\mathbf{P}_{s,i}) \;+\; \mathbf{b}_p
\;\in\; \mathbb{R}^{d},
\]
with model width $d$ (e.g., $d{=}384$ or $512$). This preserves local lead-time morphology (e.g., ST and T dynamics) while controlling sequence length.

\subsection{Positional, Lead, and Stage Embeddings}
We inject domain knowledge via additive embeddings:
\begin{align*}
\tilde{\mathbf{t}}_{s,i} \;=\;
\mathbf{t}_{s,i}
\;+\; \mathbf{e}^{\text{pos}}_{s,i}
\;+\; \mathbf{e}^{\text{lead}}_{s,i}
\;+\; \mathbf{e}^{\text{stage}}_{s},
\end{align*}
where $\mathbf{e}^{\text{pos}}_{s,i}$ encodes intra–stage patch position (2D or separable 1D positional encodings along lead and time), $\mathbf{e}^{\text{lead}}_{s,i}$ encodes which leads are included in the patch (learned embedding that can softly favor \{II, III, aVF, V5, V6\}), and $\mathbf{e}^{\text{stage}}_{s}$ encodes the exercise stage/MET level (learned lookup or small MLP over the numeric MET). \textbf{No phase/segment embedding is used.}

\subsection{Hierarchical Encoder}
We adopt a two-level transformer that first summarizes each stage locally and then reasons globally across stages.

\paragraph*{Stage (local) encoder.}
For each stage $s$, we prepend a learned token $\mathbf{c}^{\text{stg}}$ and process
\[
\mathbf{Z}^{\text{local}}_{s}\;=\;\text{Transformer}_{\text{local}}
\big([\mathbf{c}^{\text{stg}},\,\tilde{\mathbf{t}}_{s,1},\ldots,\tilde{\mathbf{t}}_{s,N_s}]\big)\in\mathbb{R}^{(N_s+1)\times d}.
\]
The stage summary is the output \emph{CLS}-like vector $\mathbf{z}_s=\mathbf{Z}^{\text{local}}_{s}[0]\in\mathbb{R}^{d}$. Depth $D_{\!l}{=}2{\sim}4$, heads $H_{\!l}{=}6$, MLP ratio $r{=}4$ are effective defaults.

\paragraph*{Cross-stage (global) encoder.}
We concatenate stage summaries with a global \emph{CLS} token $\mathbf{c}^{\text{glob}}$:
\[
\mathbf{Z}^{\text{glob}}\;=\;\text{Transformer}_{\text{global}}
\big([\mathbf{c}^{\text{glob}},\,\mathbf{z}_1,\ldots,\mathbf{z}_S]\big)\in\mathbb{R}^{(S+1)\times d},
\]
and use $\mathbf{g}=\mathbf{Z}^{\text{glob}}[0]\in\mathbb{R}^{d}$ as the patient-level representation. Typical settings: $D_{\!g}{=}4{\sim}6$, $H_{\!g}{=}6{\sim}8$, MLP ratio $r{=}4$.

\subsection{Classifier Heads and Objectives}
The primary CAD head maps $\mathbf{g}$ to logits $\boldsymbol{\ell}\in\mathbb{R}^{2}$:
\[
\boldsymbol{\ell}\;=\;\mathbf{W}_c\,\mathbf{g}+\mathbf{b}_c,\qquad
\hat{y}\;=\;\text{softmax}(\boldsymbol{\ell}),
\]
trained with cross-entropy. For class imbalance, focal loss or class-weighted CE can be used. An optional auxiliary \emph{stage-risk} head $\phi(\mathbf{z}_s)$ predicts per-stage risk; patient risk can be aggregated (e.g., attention-weighted average or max) to encourage learning of peak-exercise cues.

\subsection{Training Protocol}
We use AdamW with learning rate $2{\times}10^{-4}$, weight decay $5{\times}10^{-2}$, cosine decay with 5-epoch warmup, batch size 8–16 patients (gradient accumulation as needed), $60$–$100$ epochs. Regularization: dropout $0.1$, stochastic depth $0.1{\sim}0.2$, label smoothing $0.05$. Data augmentation includes mild Gaussian noise, low-frequency baseline drift, time-warp, and gain jitter, avoiding morphology distortion.

\subsection{Inference and Calibration}
Given $\mathbf{X}$ we compute $\hat{y}$ from $\mathbf{g}$. When using stage-risk, we combine global \emph{CLS} and aggregated stage risks (e.g., arithmetic mean or learned attention). Temperature scaling on a validation split improves calibration. We report sensitivity, specificity, accuracy, PPV/NPV, and AUC against ICA-defined ground truth.

\subsection{Complexity Considerations}
Patch size $(p_h,p_w)$ trades token count for locality; hierarchical encoding reduces global sequence length from $\sum_s N_s$ to $S{+}1$ at the second level, enabling longer $\mathbf{X}$ and more stages without quadratic cost explosion. Lead and stage embeddings add $\mathcal{O}(d)$ parameters and negligible FLOPs.

\noindent{\bf DL + EKE execution time:} Quantification of computation complexity of any machine with deep learning is difficult since it involves several hyper-parameter training. However, Table \ref{tab1} shows that DL+EKE is not the fastest nor is it the slowest. The fastest is knowledge extraction since it does not involve DL. Pretrained CNN is faster than DL+EKE. However, DL+EKE is faster than LVM and ViT approaches.

\section{A5: DR specifics}

\noindent{\bf DR grading} \noindent{\bf benchmarks:} We use the Eyepacs~\cite{Cuadros2009EyePACS}, Aptos~\cite{APTOS2019BlindnessDetection}, Messidor 1~\cite{Decenciere2014Messidor}, and Messidor 2~\cite{Messidor2Consortium2024} datasets.

\noindent{\bf DR Baselines:} We use the latest baselines reported in~\cite{galappaththige2024generalizing,banerjee2026wacv} described in Appendix. that includes DRGen~\cite{atwany2022drgen}, SD-ViT~\cite{sultana2022sd_vit}, SPSD-ViT~\cite{galappaththige2024spsd}, and ERM-ViT~\cite{teterwak2025erm++}. We also used latest VLM models such as CLIP~\cite{radford2021clip}, ordinal CLIP~\cite{li2022ordinalclip}, and VLM specifically trained on DR images, CLIP-DR~\cite{yu2024clipdr}.

\begin{table}[h]
  \centering
  \caption{Clinical Signs of DR and Their Diagnostic Significance}
  \label{tab:dr_symptoms}
  \footnotesize
  \begin{tabularx}{\columnwidth}{@{}lX@{}}
    \toprule
    \textbf{Symptom} & \textbf{Key Observations and Diagnostic Relevance} \\
    \midrule
    Microaneurysms & Tiny red capillary dilations in the retina; earliest sign of mild NPDR. Their progression correlates with disease severity~\cite{bhavsar2019diabetic}. \\
    \addlinespace
    Hemorrhages & Includes dark red dot/blot and flame-shaped types indicating microvascular leakage. Severe NPDR is marked by more than 20 hemorrhages in all quadrants; risk of PDR rises to 50\% within a year~\cite{wong2018guidelines}. \\
    \addlinespace
    Hard Exudates & Sharp yellow lipid-rich deposits from chronic leakage, often in/near the macula. Indicative of risk for diabetic macular edema (DME), a major cause of vision loss~\cite{etdrs1991grading}. \\
    \addlinespace
    Cotton Wool Spots & Fluffy white retinal lesions caused by nerve-fiber-layer infarctions. Signify retinal ischemia in moderate to severe NPDR~\cite{diabetic_retinopathy_statpearls}. \\
    \addlinespace
    Subhyaloid Hemorrhages & Boat- or D-shaped hemorrhages between the retina and the hyaloid face, typically from ruptured neovascular vessels. Hallmark of proliferative DR~\cite{campbell1995subhyaloid}. \\
    \addlinespace
    Neovascularization & Fragile new vessel growth on the optic disc (NVD) or elsewhere on the retina (NVE). Defining trait of PDR; untreated cases face ~60\% vision loss within five years~\cite{dandona2001pdr}. \\
    \bottomrule
  \end{tabularx}
\end{table}

\section{A6: Data description}
Resting state fMRI data was acquired from 2 independent centers: Centre 1 and Centre 2 using IRB process and cross-university agreement. Center 1 dataset consists of 52 patients, age 3 months – 18 years (all children), 23 Male and 29 Female. The MRI images were acquired using a 3T MRI, Ingenuity Philips Medical system with a 32-channel head coil. The rs-fMRI parameters were set at TR 2000ms, TE 30 ms, matrix size 80 x 80, flip angle 80o, number of slices 46, slice thickness 3.4 mm with no gap, in-plane resolution 3x3 mm, interleaved acquisition, and number of total volumes 600, in two 10-min runs, with total time of 20 mins. Centre 2 dataset consists of 31 patients with ages spanning from 2 months to 62 years (20 childnre, 11 adults), 14 Male and 17 Female. The MRI images were acquired using Siemens’s MAGNETOM Prisma FIT scanner, with the following parameters configured: TR 2010ms, TE 32ms, flip angle 82o, slice thickness 4mm and spacing between slices 4.

\section{A7: Hyperparameter settings} We used Keras Hyperband tuner algorithm with objective of minimizing validation loss to obtain the hyperparameters of DL part. The 2D CNN architecture was used as it gave the best results for noise IC classification (Table I). CNN’s hyperparameters were fine-tuned using 80\% training data, reserving the remaining 20

i)	Across\_trial cross validation (Center A): The final hyperparameters were:
a)	Number of convolutional layers: 3
b)	Number of 3 X 3 filters in convolutional layer 1, 2 and 3: 64, 64 and 256 respectively.
c)	Number of neurons in dense fully connected layer: 704.
d)	Learning rate: 0.0001.
e)	Dropout rate: 0.33.
f)	Batch\_size = 32.

ii)	Across\_trial cross validation (Center B): The final hyperparameters were:
a)	Number of convolutional layers: 6
b)	Number of 3 X 3 filters in convolutional layer 1, 2, 3, 4, 5 and 6: 128, 16,64, 512, 512, and 256 respectively.
c)	Number of neurons in dense fully connected layer: 3008.
d)	Learning rate: 0.0001.
e)	Dropout rate: 0.33.
f)	Batch\_size = 32.

iii)	Aggregated Trial validation: 
a)	Number of convolutional layers: 2
b)	Number of 3 X 3 filters in convolutional layer 1 and 2: 16 and 64 respectively.
c)	Number of neurons in dense fully connected layer: 1472.
d)	Learning rate: 0.0001.
e)	Dropout rate: 0.33.
f)	Batch\_size = 16.

\begin{algorithm}
\caption{RareSaGe Algorithm}

\scriptsize

\textbf{Input:} Raw data $Y$ with $n$ classes $\{C_1, C_2, \dots, C_n\}$, $\theta_M, \sigma_\theta$, confidence threshold $t_c$, set of classifiers $\mathcal{M_{DL}}$ and $\mathcal{M_{QO}}$.

\begin{algorithmic} [1]
\STATE Sample set $\Psi = Y$
\WHILE{$\Psi$ is not empty and significant change in validation accuracy}
\STATE Compute proportion $\theta_i$ of each class $C_i$, where $i \in \{1, 2, \dots, n\}$
\STATE Identify rare class $C_{rare}$ such that $\theta_{rare}$ satisfies Definition \ref{def:Rare}.
\STATE Compute the overlap class $c_{o}$ utilizing the highest similarity with $C_{rare}$ in a class agnostic embedding space representation.

\STATE Divide the dataset into  $c_o$ and $\neg c_o$ classes

\FOR{ each classifier $M_d \in  \mathcal{M_{DL}}$}
\STATE Compute $\theta(M_d)$ on the set $\Psi$ for $c_{o}$ and $\neg c_{o}$ 
\ENDFOR

\STATE Choose classifier with minimum $\theta$: $M_d \leftarrow \arg\min_{M_d} \theta(M_d)$. 

\STATE Divide $\neg c_o$ into $c_r$ and $\neg c_r$
\FOR{ each classifier $M_d \in  \mathcal{M_{QO}}$}
\STATE Compute $\theta(M_d)$ on the set $\Psi$ for $c_{r}$ and $\neg c_{r}$

\ENDFOR
    
\STATE Choose classifier with minimum $\theta$: $M_d \leftarrow \arg\min_{M_d} \theta(M_d)$.

\STATE Compute confidence scores for classifiers $M_{DL}$ and $M_{QO}$
   \STATE Choose classifier $M_{QO}$ with score $> t_c$ for rare class labelling

\STATE \textbf{Return} Final labels: Overlap Class, Rare Class, Non-rare Class
\ENDWHILE
\end{algorithmic}
\end{algorithm}

\section{A8: Application of RareSaGe in other examples}
We show the exact details for stress ECG based CAD and DR grading approach.
\subsection{Stress ECG based CAD detection}
The overall architecture is shown in Fig. \ref{transformerModel}.
\subsubsection{Transitioning from Spatial to Temporal}\label{AA}

The ECG images were transformed into time-series signals, each standardized to a length of 493. Signals with lower sampling rates were upsampled, while shorter ones were padded with -10. This transformation involved using optical character recognition (OCR) software to extract textual information and pixel data. Preprocessing removed textual artifacts and isolated the ECG signal using color-based edge detection. The resulting signal was normalized within the range [-1.0, 1.0]. Manual validation excluded 65 patients from the dataset where the converted signal deviated from the original stress ECG image.

While Vision Transformers (ViTs) excel in learning spatial features from images, they struggle with domain-specific features crucial for CAD diagnosis. Converting ECG images into time series data offered several advantages. It enabled the utilization of temporal features crucial for diagnosing cardiac abnormalities, improving CAD detection accuracy. Expert knowledge integration refined the model's focus on clinically significant information. Additionally, artifact removal and noise reduction techniques mitigated false positives.

\subsubsection{Expert Knowledge}

 CAD disrupts cell membranes through ion channel alterations, leading to cellular dysfunction. Stress ECG targets ST segment depression, a key ischemic heart disease criterion, with a 1 mm threshold at 80 milliseconds from the J point, improving CAD detection PPV. Depolarization changes, less studied in stress ECG, also influence CAD detection. QRS complex changes during ischemia are limited due to composite signals and eye resolution constraints. However, longer repolarization duration improves surface ECG representation despite being visually subtle. Inferior (LII, LIII, aVF) and lateral (V5, V6) leads are critical for detecting ischemic changes. Repolarization-induced ST depression is non-localizing. CAD prognosis is impacted by MET, assessed via the Duke treadmill score. Time-series data from five leads are aligned to represent the highest MET ECG signal per patient, with 493 time steps analyzed by the temporal transformer.

 \subsubsection{Integration of expert knowledge in transformer architecture}
The expert knowledge is integrated by adding knowledge guided masks on the raw data. In specific we use two types of masks: \textbf{a) speed mask}, is a one hot encoding of METs that should be used according to experts and determines the speeds $S$, and \textbf{b) lead mask}, is a one hot encoding of the leads that should be used as guided by experts and determines $L$.
\subsubsection{Time-series Encoder-Encoder Transformer}
\begin{figure}[]
\centerline{\includegraphics[width=0.5\textwidth, height=0.4\textheight]{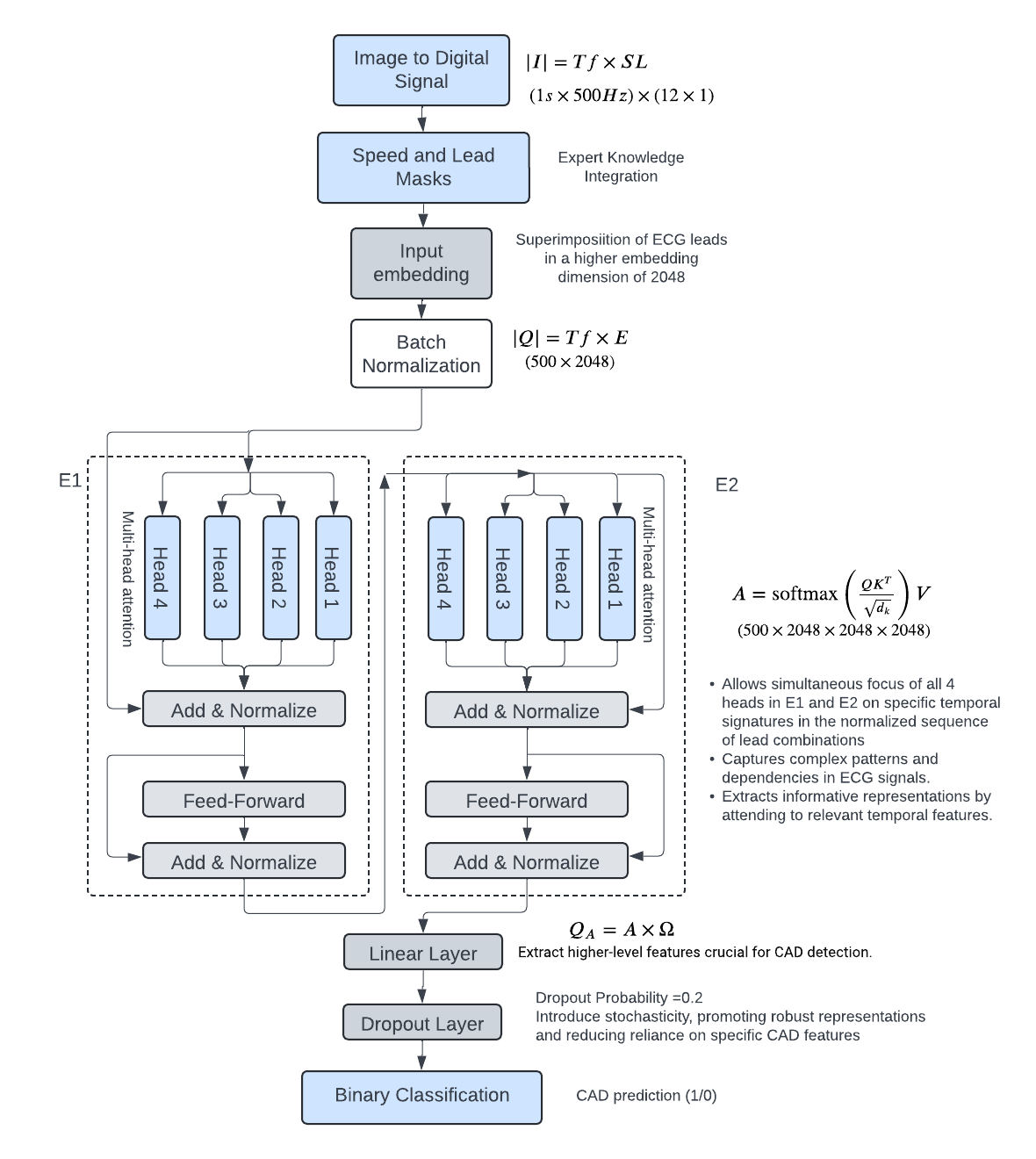}}
\caption{Expert-Guided Transformer: This 2 Encoder (E1-E2)  model incorporates expert knowledge on METs and lead selection into the input stage and integrates CAD temporal characteristics into the transformer's attention layer.}
\label{transformerModel}
\end{figure}

The model first converts raw ECG data $X$ ($493\times5$) into input embeddings using learnable weights $W$ ($5\times2048$) and bias $b$. ReLU activation adds non-linearity, capturing temporal patterns while subsequent batch normalization standardizes signal values across mini-batches during training. It was trained using binary cross-entropy (BCE) loss function for classification. The Query(Q), Key(K), and Value(V) vectors are derived from this sequence, allowing the model to capture relationships between different elements within the sequence itself. The self-attention mechanism of each head assigns weights (\(W^Q\) [2048x\(d_q\)], \(W^K\) [2048x\(d_k\)], \(W^V\) [2048x\(d_v\)]) to each element based on its relevance to other lead combination signal values in the sequence. This is achieved through a process of computing the similarity between the Q and K vectors, followed by scaling and applying a softmax function to obtain attention scores. 
\[
\text{Q} = XW^Q, \quad \text{K} = XW^K, \quad \text{V} = XW^V
\]

where \(X\) is the input sequence of dimension [500x2048], and \(W^Q\), \(W^K\), and \(W^V\) are learnable weight matrices.

The attention score \(A\) is computed as:
\[
A = \text{softmax} \left( \frac{QK^T}{\sqrt{d_k}} \right) V
\]
where \(d_k\) is the dimension of the key vectors.
\[
\text{MultiHead}(Q, K, V) = \text{Concat}(\text{h}_1,\text{h}_2 , \text{h}_3, \text{h}_4)W^O
\]
Here, the multi-head attention is computed where \(\text{h}_1, \text{h}_2 , \text{h}_3, \text{h}_4\) represents the 4 attention heads and \(W^O\) is the output weight matrix.
These scores are then used to compute a weighted sum of the V vectors, producing the attended output.
Self-attention in the transformer model captures key temporal features while preserving context, with each of the four attention heads identifying distinct patterns and dependencies concurrently. Each head projects the sequence into distinct subspaces, enabling extraction of a wide range of features. After passing through linear transformations and ReLU activation functions, the output undergoes further processing with linear layers after multi-head attention. Dropout layers prevent overfitting by deactivating neurons during training, encouraging robust learning as in Fig.~\ref{transformerModel}.

\subsection{RareSaGe application on DR Grading}
We evaluated the proposed framework on the task of DR classification into five classes using retinal fundus images (Fig. \ref{DRFig}). Fundus images can be classified into five classes: 0, no DR, 1 mild DR, 2 Moderate DR, 3 severe, and 4 proliferative DR. Here three classes viz. 2, 3, 4 are rare classes, white 0 and 1 are not rare.  

\subsubsection{Knowledge extraction}
Domain-specific diagnostic rules for each class were curated from ophthalmological guidelines (see Table~\ref{tab:dr_symptoms} in appendix) and operationalized via automated feature extraction pipelines built using two open-source tools: YOLOv12 and a retinal vessel segmentation model (Fig. \ref{DRFig}).

For lesion-level localization, we employed the YOLOv12 object detection model~\cite{tian2025yolov12attentioncentricrealtimeobject}, a SOTA one-stage detector known for its efficiency and precision in dense object environments. YOLOv12 extends the YOLOv5/YOLOv7 series with advanced improvements including C3K2 blocks for enhanced feature extraction, Spatial Pyramid Pooling Fast (SPPF) modules for multi-scale processing, and C2PSA (Cross-Stage Partial Self-Attention) mechanisms that selectively apply attention to critical image regions~\cite{tian2025yolov12attentioncentricrealtimeobject}. The model architecture employs a modified CSPDarkNet backbone with decoupled detection heads that independently process objectness, classification, and regression tasks, achieving superior mean average precision (mAP) with real-time inference capabilities~\cite{wang2023comprehensive}. We trained YOLOv12 to detect clinically relevant lesions such as hemorrhages, hard exudates, and cotton wool spots. Bounding boxes produced by the model were post-processed and validated using Intersection over Union (IoU) scores against expert-labeled fundus images, ensuring medical fidelity. Each image passed through Yolov12 model results in a maximum 14 dimensional real vector consisting of the IoU scores.  

\subsubsection{DL model}  
\begin{figure}
\centering
\includegraphics[width=0.5\textwidth]{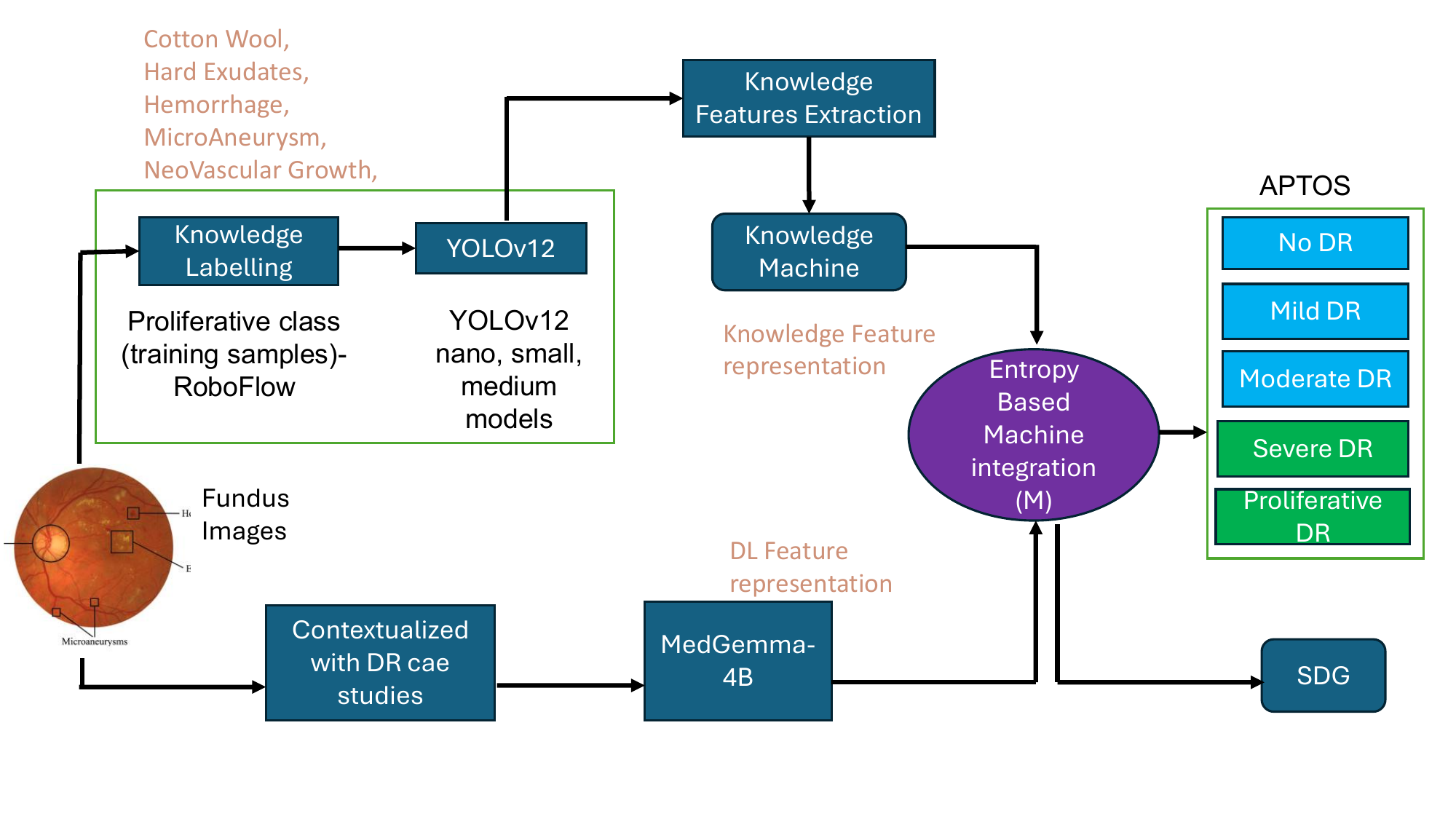}
\caption{Use of RareSaGe on Diabetes Retinopathy grading} 
\label{DRFig}
\end{figure}

Our framework leverages MedGemma-4B~\cite{medgemma_card_2025}, a 4-billion parameter vision-language model specifically pre-trained on extensive medical image-text pairs. Unlike general-purpose vision-language models that struggle with clinical terminology and diagnostic reasoning~\cite{radford2021clip,zhang2023biomedclip,li2023llava_med}, MedGemma-4B incorporates medical knowledge through its training on clinical reports (Fig. \ref{DRFig}).

\subsubsection{Knowledge DL integration}

We utilize the same CLIP based cross entropy to determine the rare class that should be separated first. Lest assume Class 4 is the rarest as observed from the cross entropy computation through CLIP. We then provide knowledge guided prompts to the MedGemma-4B model for final classification of rare class. We utilize the following prompt structure that incorporates knowledge:

\begin{scriptsize}
\texttt{``You are an ophthalmologist with 15 years of experience in diabetic retinopathy screening. Analyze this fundus image systematically, \textbf{examining the macula, optic disc, and four quadrants for: microaneurysms, hemorrhages, hard/soft exudates, venous abnormalities, intraretinal microvascular abnormalities (IRMA), and neovascularization.} According to knowledge model the knowledge vector for this fundus image is \{ 14 element vector from YOLOv12\}
Classify according to international standards \{not 4,4\}. Return only the severity number.''} 
\end{scriptsize}

After identification of Class 4 then we apply the same technique this time for the next rare class. We observe that different datasets result in different rare class sequences.

\section{A9: All DR Tables}
\noindent{\bf DR grading insights:} An interesting factor is that for multiple rare class scenarios, the sequence at which rare class is handled changes based on datasets. For example in Eyepacs and Messidor 1, the largest dataset, we handled rare class with the sequence, 4, 3, 2, 1, 0. However, for Aptos the cross entropy technique found the sequence 3, 2, 4, 1, 0. In Messidor 2 the cross entropy technique processes rare classes in the order 3, 2, 1, 0. Rightfully so because the Class 4 is absent in the dataset. The difference between Eyepacs and Aptos is interesting because in Aptos we observed that grading 4 data had a distinct laser scarring artifact. This decreased the class entropy because it is a clear indication of multiple scans correlating with higher level DR. Hence Class 4 was easily identifiable in Aptos and was not the rarest.  
\begin{table}[!htbp]
  \centering
  \caption{Across trial Training on EyePACS. Accuracy }
  \label{tab:eyepacs_source}
  \scriptsize
    \begin{tabular}{@{}p{1.4 in}@{}p{0.45 in}@{}p{0.45 in}@{}p{0.45 in}@{}p{0.35 in}@{}}
      \hline
      \textbf{Method}                  & \textbf{APTOS}      & \textbf{Messidor}   & \textbf{Messidor2} & \textbf{Average} \\
      \hline
      DRGen \cite{atwany2022drgen}     & 61.3$\pm$1.9        & \underline{54.6$\pm$1.5}        & \underline{65.4$\pm$0.1}        & 60.4             \\
      ERM-ViT \cite{teterwak2025erm++} & 69.1$\pm$1.4       & 50.4$\pm$0.3        & 62.8$\pm$0.2        & 60.8             \\
      SD-ViT \cite{sultana2022sd_vit}  & 69.3$\pm$0.3        & 50.0$\pm$0.5        & 62.9$\pm$0.2       & 60.7             \\
      SPSD-ViT \cite{galappaththige2024spsd}
                                        & \textbf{75.1$\pm$0.5} & 50.5$\pm$0.8        & 62.2$\pm$0.4        & \underline{62.5}             \\
      DL+EKE                    & \underline{73.2$\pm$ 0.4}                & \textbf{69.5$\pm$ 0.8}       & \textbf{80.5$\pm$ 0.4}       & \textbf{74.4}    \\
      \hline
    \end{tabular}
\end{table}

\begin{table}[!htbp]
\centering
\caption{Across trial: Training on Aptos. Accuracy }
\label{tab:aptos_source}
\scriptsize
\begin{tabular}{@{}p{1.4 in}@{}p{0.45 in}@{}p{0.45 in}@{}p{0.45 in}@{}p{0.35 in}@{}}
\hline
\textbf{Method}                  & \textbf{EyePACS}  & \textbf{Messidor} & \textbf{Messidor2} & \textbf{Average} \\
\hline
DRGen \cite{atwany2022drgen}     & 67.5$\pm$1.8      & \underline{46.7$\pm$0.1}      & \underline{61.0$\pm$0.1}     & 58.4             \\
ERM-ViT \cite{teterwak2025erm++} & 67.8$\pm$1.4      & 45.5$\pm$0.2      & 58.8$\pm$0.4       & 57.3             \\
SD-ViT \cite{sultana2022sd_vit}  & \underline{72.0$\pm$0.8}      & 45.4$\pm$0.1      & 58.5$\pm$0.2       & \underline{58.6}             \\
SPSD-ViT \cite{galappaththige2024spsd} & 71.4$\pm$0.8 & 45.6$\pm$0.1      & 58.8$\pm$0.2       & \underline{58.6}           \\
{DL+EKE }          & \textbf{77.8$\pm$0.8}     & \textbf{49.0$\pm$0.2}     & \textbf{64.0$\pm$0.2}      & \textbf{63.6}    \\
\hline
\end{tabular}
\end{table}

\begin{table}[!htbp]
\centering
\caption{Across trial: Training on Messidor. Accuracy }
\label{tab:messidor1_source}
\scriptsize
  \setlength{\tabcolsep}{4pt}
\begin{tabular}{@{}p{1.25 in}cccc@{}}
\hline
\textbf{Method}                  & \textbf{APTOS}    & \textbf{EyePACS}  & \textbf{Messidor-2} & \textbf{Average} \\
\hline
DRGen \cite{atwany2022drgen}     & 41.7$\pm$4.3      & 43.1$\pm$7.9      & 44.8$\pm$0.9       & 43.2             \\
ERM-ViT \cite{teterwak2025erm++} & 45.3$\pm$1.3     & 52.4$\pm$3.2      & 58.2$\pm$3.2      & 51.9             \\
SD-ViT \cite{sultana2022sd_vit}  & 44.3$\pm$0.9      & 53.2$\pm$1.6      & 57.8$\pm$2.4       & 51.7             \\
SPSD-ViT \cite{galappaththige2024spsd} & \underline{48.3$\pm$1.1} & \underline{57.4$\pm$2.1}      & \underline{62.2$\pm$1.6}       & \underline{55.9}             \\
{DL+EKE}          & \textbf{56.0$\pm$0.8}     & \textbf{80.0$\pm$2.1}     & \textbf{65.2$\pm$2.4}      & \textbf{67.1}    \\
\hline
\end{tabular}
\end{table}

\begin{table}[!htbp]
\centering
\caption{Across trial: Training on Messidor-2. Accuracy }
\label{tab:messidor2_source}
\scriptsize
\begin{tabular}{@{}p{1.6 in}@{}p{0.45 in}@{}p{0.45 in}@{}p{0.45 in}@{}p{0.35 in}@{}}
\hline
\textbf{Method}                  & \textbf{APTOS}    & \textbf{EyePACS}  & \textbf{Messidor}   & \textbf{Average} \\
\hline
DRGen \cite{atwany2022drgen}     & 40.9$\pm$3.9      & 69.3$\pm$1.0      & 61.3$\pm$0.8       & 57.7             \\
ERM-ViT \cite{teterwak2025erm++} & 47.9$\pm$2.1      & 67.4$\pm$0.9      & 59.6$\pm$3.9       & 58.3             \\
SD-ViT \cite{sultana2022sd_vit}  & 51.8$\pm$0.9     & 68.7$\pm$0.6      & \underline{62.0$\pm$1.7}       & 60.8             \\
SPSD-ViT \cite{galappaththige2024spsd} & \underline{52.8$\pm$2.0} & \underline{72.5$\pm$0.3}      & 61.0$\pm$0.8       & \underline{62.1 }            \\
{DL+EKE }          & \textbf{69.7$\pm$1.8}     & \textbf{77.8$\pm$0.3}     & \textbf{67.7$\pm$0.8}      & \textbf{71.7}    \\
\hline
\end{tabular}
\end{table}

\begin{table}[!htbp]
\centering
\caption{Aggregate trial Performance for DR.}
\label{tab:vlm_performance}
\scriptsize
\begin{tabular}{l|ccc}
\hline
\textbf{VLM Method} & \textbf{APTOS} & \textbf{Messidor} & \textbf{Average} \\
\hline
\multicolumn{4}{c}{\textit{F1-Score Performance (\%)}} \\
\hline
CLIP~\cite{radford2021clip}          & 44.3 & 39.6 & 41.9 \\
OrdinalCLIP~\cite{li2022ordinalclip} & 45.7 & 41.8 & 43.8 \\
CLIP-DR~\cite{yu2024clipdr}          & 46.3 & 47.3 & 46.8 \\
{DL+EKE}          & \textbf{72.0} & \textbf{78.2} & \textbf{75.1} \\\hline
\end{tabular}
\end{table}

\begin{table}[h!]
\scriptsize
\caption{Aggregate-Trial Validation Results for SOZ.}
\begin{center}
\begin{tabular}{|@{}p{0.35 in}@{}|p{0.5 in}@{}|p{0.5 in}@{}|p{0.5 in}|p{0.5 in}@{}|p{0.4 in}@{}|}
\hline
\textbf{Repeat} &\textbf{Accuracy} &\textbf{Precision} &\textbf{Sensitivity} &\textbf{F1 score}  &\textbf{Mean F1} \\ 
    \hline
     Repeat 1 & 91.6\% (6.9) & 92.8\% (7.8) & 98.7\% (2.8) &  95.4\% (3.8)&  94.7\%   \\ 
    \cline{1-5}
      Repeat 2  &89.1\% (6.8)     & 92.5\% (5.0)    & 95.8\% (3.7)         &  94.1\% (3.7)       & (3.7)\\ 
     \cline{1-5}
     Repeat 3  &90.3\% (7.0)  &  92.6\% (5.0)       & 97.2\% (3.8)         &  94.8\% (3.8)       &  \\ 
    \cline{1-6}
  \end{tabular}
  \label{table2}
\end{center}
\end{table}

\begin{table*}
\scriptsize
\centering
\caption{Performance results of across-trial validation—single domain generalizability for rare class detection.}\label{tab2}
\begin{tabular}{lllllllll|} 
\hline
\multicolumn{2}{|l|}{\textbf{Method}} & \textbf{Accuracy} & \textbf{Precision} & \textbf{Sensitivity} & \textbf{F1-score} & \textbf{Average F1-score} & \textbf{Time(min)} & \textbf{Ablation}\\
\hline
\multicolumn{2}{|l|}{Pre-trained CNN Train A, Test B}            & 67.7\% & 87.5\% & 75.0\% & 80.7\% & 49.1\% & 28(6) & DL only \\
\multicolumn{2}{|l|}{Pre-trained CNN Train B, Test A}            &  9.6\% & 62.5\% & 10.2\% & 17.5\% &         & 32(6) & DL only \\ \hline

\multicolumn{2}{|l|}{Pre-trained ViT small Train A, Test B}       & 64.5\% & 86.9\% & 71.4\% & 78.4\% & 77.2\% & 45(12) & DL only \\
\multicolumn{2}{|l|}{Pre-trained ViT small Train B, Test A}       & 61.5\% & 91.4\% & 65.3\% & 76.1\% &         & 42(13) & DL only \\ \hline

\multicolumn{2}{|l|}{Pre-trained ViT base Train A, Test B}        & 45.1\% & 82.3\% & 50.0\% & 62.2\% & 60.8\% & 55(10) & DL only \\
\multicolumn{2}{|l|}{Pre-trained ViT base Train B, Test A}        & 42.3\% & 88.0\% & 44.8\% & 59.4\% &         & 50(8)  & DL only \\ \hline

\multicolumn{2}{|l|}{ViT trained from scratch, Train A, Test B}   & 12.9\% & 57.1\% & 14.2\% & 22.7\% & 46.3\% & 60(15) & DL only \\
\multicolumn{2}{|l|}{ViT trained from scratch, Train B, Test A}   & 53.8\% & 90.3\% & 57.1\% & 70.0\% &         & 58(18) & DL only \\ \hline

\multicolumn{2}{|l|}{LVM fine tuned—contrastive loss, Train A, Test B} & 64.5\% & 86.9\% & 71.4\% & 78.4\% & 46.3\% & 52(14) & DL only \\
\multicolumn{2}{|l|}{LVM fine tuned—contrastive loss, Train B, Test A} &  7.6\% & 57.1\% &  8.1\% & 14.2\% &         & 48(11) & DL only \\ \hline

\multicolumn{2}{|l|}{LVM fine tuned—cross entropy loss, Train A, Test B} & 51.6\% & 84.2\% & 57.1\% & 67.9\% & 37.6\% & 47(13) & DL only \\
\multicolumn{2}{|l|}{LVM fine tuned—cross entropy loss, Train B, Test A} &  3.8\% & 40.0\% &  4.0\% &  7.3\% &         & 43(16) & DL only \\ \hline

\multicolumn{2}{|l|}{Knowledge based system, Train A, Test B}     & 83.8\% & 89.6\% & 92.8\% & 91.2\% & 78.9\% & 13(6)  & Knowledge only \\
\multicolumn{2}{|l|}{Knowledge based system, Train B, Test A}     & 50.0\% & 89.6\% & 53.0\% & 66.6\% &         & 12(6)  & Knowledge only \\ \hline

\multicolumn{2}{|l|}{\textbf{DL+EKE Train A, Test B}}             & \textbf{90.3\%} & \textbf{90.3\%} & \textbf{100\%} & \textbf{94.9\%} & \textbf{90.2\%} & \textbf{35(10)} & \textbf{DL+EKE} \\
\multicolumn{2}{|l|}{\textbf{DL+EKE Train B, Test A}}             & \textbf{75.0\%} & \textbf{92.8\%} & \textbf{79.5\%} & \textbf{85.6\%} &         & \textbf{38(9)}  & \textbf{DL+EKE} \\ \hline
\multicolumn{2}{|l|}{\textbf{DL+EKE Train A Children, Test B Adults}} & \textbf{90.9\%} & \textbf{90.9\%} & \textbf{100\%} & \textbf{95.2\%} & \textbf{95.2\%} & \textbf{36(7)}  & \textbf{DL+EKE} \\
\hline
\end{tabular}
\end{table*}

\section{A10: Code Availability}
https://github.com/ImpactLabASU/CADDiagnosis

\section{Acknowledgments}
We thank Payal Kamboj, Riya Sudhakar Salian, Midhat Urooj, and Kuntal Thakur for data collection, and results compilation on SOZ, CAD, and DR case studies. This project is partially funded by NSF FDTBiotech grant (2436801), NIH R21 grant (1R21HL175632), Mayo Cardiovascular research grant, Mayo ASU seed grant, and Arizona New Economy Initiative grant. We thank Phoenix Children Hospital and UNC Chapel Hill for Epilepsy data and Dr. Boerwinkle for sharing expert knowledge on SOZ.

\end{document}